\documentclass[12pt,preprint]{aastex}

\slugcomment{Accepted for publication in the Astrophysical Journal Letters}
\shorttitle{Complex Radial Moctions in Barnard 68}
\shortauthors{Maret, Bergin and Lada}

\begin{document}

\title{Using Chemistry to Unveil the Kinematics of Starless Cores:
  Complex Radial Motions in Barnard 68}

\author{S\'ebastien Maret, Edwin A. Bergin}
\affil{Department of Astronomy, University of Michigan, 500 Church
  Street, Ann Arbor, MI 48109-1042, USA}
\and
\author{Charles J. Lada}
\affil{Harvard-Smithsonian Center for Astrophysics, 60 Garden Street,
    Cambridge, MA 02138, USA}

\begin{abstract}
  We present observations of $^{13}$CO, C$^{18}$O, HCO$^{+}$,
  H$^{13}$CO$^{+}$, DCO$^{+}$ and N$_{2}$H$^{+}$ line emission towards
  the Barnard 68 starless core. The line profiles are interpreted
  using a chemical network coupled with a radiative transfer code in
  order to reconstruct the radial velocity profile of the core. Our
  observations and modeling indicate the presence of complex radial
  motions, with the inward motions in the outer layers of the core but
  outward motions in the inner part, suggesting radial
  oscillations. The presence of such oscillation would imply that B68
  is relatively old, typically one order of magnitude older than the
  age inferred from its chemical evolution and statistical core
  lifetimes. Our study demonstrates that chemistry can be used as a
  tool to constrain the radial velocity profiles of starless cores.
\end{abstract}


\section{Introduction}
\label{sec:introduction}

Starless cores likely represent the earliest stage of the formation of
a star. In the standard view of star formation, cores form out an
initially magnetically sub-critical cloud, and evolve in a
quasi-static fashion through ambipolar diffusion
\citep{Shu87,Mouschiovas91}. In the opposite picture, cores are
dynamic objects that form by shocks in a turbulent flow, fragment and
collapse (or disappear) in a few crossing times
\citep{Padoan01,Ballesteros-Paredes2003}. Velocity measurements can
provide important constraints on these two scenarios. For example,
\cite{Tafalla98} examined the kinematics of L1544 and inferred infall
velocities up to 0.1 km s$^{-1}$ \emph{a priori} incompatible with
sub-critical ambipolar diffusion models. However, \cite{Ciolek00}
argued that these velocities can be understood if the core is
super-critical.

Line of sight velocities in cores can be established from observations
of self-absorbed line profiles. In an infalling core, if the
excitation temperature increases towards the center (as one can expect
in a centrally condensed core with a roughly constant temperature),
self-absorbed lines are expected to be asymmetric, with a blue peak
brighter than the red peak \cite[see][]{Evans99}. This technique has
been used in the past to detect collapse motions in starless cores
and protostars \cite[see][for a review]{Myers00}. One of the major
difficulties of this technique is the large degree of chemical
complexity within these objects. For example, numerous molecules are
observed to deplete in the densest regions of starless cores
\cite[e.g.][]{Bergin02a,Tafalla02}, thus hampering our ability to
measure the velocity in the dense central regions of these objects.

Nevertheless, with a detailed knowledge of the chemistry of these
objects, one can choose appropriate molecular transitions to trace
different part of the cores, and unveil their radial kinematic
structure \citep{Bergin03b,vanderTak05}. In this paper, we use line
observations of $^{13}$CO, C$^{18}$O, HCO$^{+}$, H$^{13}$CO$^{+}$,
DCO$^{+}$ and N$_{2}$H$^{+}$ to reconstruct the velocity profile of
the Barnard 68 core (hereinafter B68). With a well defined physical
\citep{Alves01,Bergin06b} and chemical
\citep{Bergin02a,Maret06,Maret07a} structure, this core is an ideal
target for such a study. For this we compare the observed line
profiles with the predictions of a chemical network coupled with a
radiative transfer code. The paper is organized as
follow. Observations are presented in \S \ref{sec:observations}. The
analysis is detailed in \S \ref{sec:analysis-results}, and the results
are discussed in \S \ref{sec:discussion}.

\section{Observations}
\label{sec:observations}

Maps of H$^{13}$CO$^{+}$ (1-0) ($\nu = 86.754288$ GHz), HCO$^{+}$
(1-0) ($\nu = 89.188523$ GHz), N$_{2}$H$^{+}$ (1-0) ($\nu = 93.173772$
GHz) C$^{18}$O (1-0) ($\nu = 109.782173$ GHz), DCO$^{+}$ (2-1) ($\nu =
144.077289$ GHz), DCO$^{+}$ (3-2) ($\nu = 216.112604$ GHz) and
HCO$^{+}$ (3-2) ($\nu= 267.557526$ GHz) transitions were observed
towards B68 ($\alpha = 17^\mathrm{h} 22^\mathrm{m} 38.2^\mathrm{s}$
and $\delta = -23 \degr 49 \arcmin 34.0 \arcsec$; J2000) in April 2002
and September 2002 with the IRAM-30m telescope.  The $^{13}$CO (2-1)
($\nu = 220.398684$ GHz) transition was observed in September 2003
with the \emph{Caltech Submillimeter Observatory} (CSO).  These
observations have already been presented in \citet{Bergin02a},
\citet{Bergin06b}, \citet{Maret06}, and \citet{Maret07a}, and we refer
the reader to these papers for technical details.

Transitions of HCO$^{+}$ (4-3) ($\nu = 356.734134$ GHz) and
N$_{2}$H$^{+}$ (3-2) ($\nu = 279.511832$ GHz) were observed in August
2005 using the \emph{Atacama Pathfinder eXperiment} telescope
(APEX). The APEX-2A receiver was used together with the FFTS
spectrometer, which provides a spectral resolution of 61 kHz.  The
telescope half power beam size is 22\arcsec\ at 280 GHz and 17\arcsec\
at 357 GHz. The data were calibrated using the chopper wheel method,
and the system temperature was $\sim$100 K at 280 GHz and 130-180 K at
357 GHz. The observations were converted on the $T_\mathrm{mb}$ scale
assuming a main beam efficiency of 70\% and a forward efficiency of
95\% (C. de Breuck, priv. comm.). These observations were obtained in
beam switching mode.

Fig. \ref{fig1} present the observations together with our model
predictions (see \S~\ref{sec:analysis-results} for a description of
the model). The lines show a variety of different profiles, from
Gaussian to self-absorbed and asymmetric double peaked
profiles. C$^{18}$O, $^{13}$CO and H$^{13}$CO$^{+}$ are probably
optically thin (or only moderately optically thick), and have
centrally peaked Gaussian profiles. On the other hand, HCO$^{+}$ (1-0)
and (3-2) lines are self-reversed and show the typical blue asymmetry
(blue peak brighter than the red peak). These line profiles suggests
that collapsing motions are present along the line of
sight. Surprisingly, self-absorbed HCO$^{+}$ (4-3), DCO$^{+}$ (3-2),
and the central component of the N$_{2}$H$^{+}$ (1-0) line exhibit red
asymmetry (red peak brighter than the blue peak), indicating
expansion. Finally, the DCO$^{+}$ (2-1) line is self-absorbed but
symmetric, hinting at the presence of static gas along the line of
sight. The absorption dip velocity of self-reversed lines is
consistent with the ${v_\mathrm{LSR}}$ of the source (3.31
$\mathrm{km\, s^{-1}}$), as determined from the $\mathrm{C^{18}O}$
(1-0) line. In the following, we model the line profiles in order to
constrain the line-of-sight velocity of the core.

\section{Analysis and results}
\label{sec:analysis-results}

\begin{figure}
  \epsscale{0.5}
  \plotone{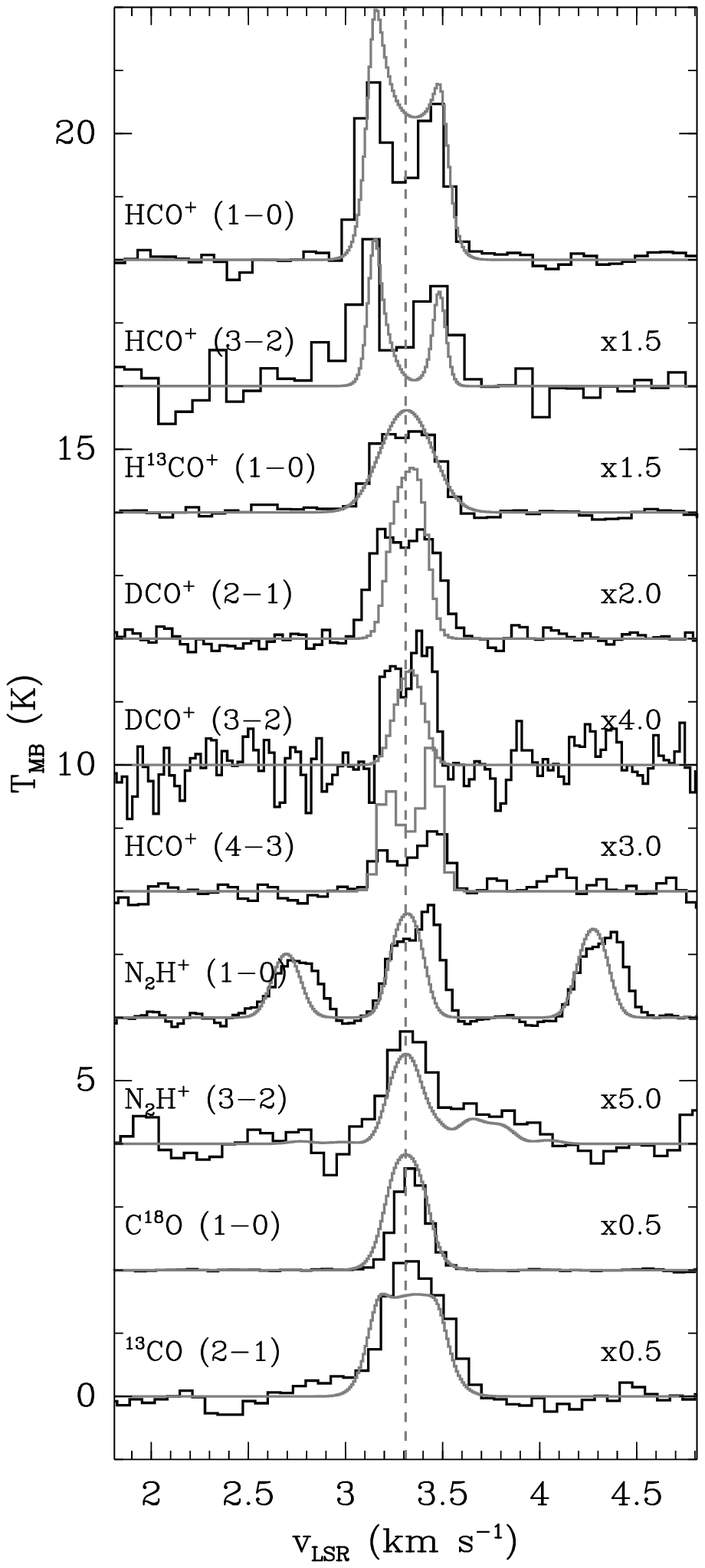}
  \caption{Comparison between the line profiles observed towards the
    extinction peak of B68 (black histograms) and the predictions of
    our ``best fit'' model (solid grey lines). For the clarity of the
    plot, lines are shifted vertically, and some of them are scaled by
    the factor mentioned on the right of each curve. \label{fig1}
  }
\end{figure}

We have modelled the spectral line profiles shown on Fig. \ref{fig1}
following the approach described in \citet{Bergin06b}, \citet{Maret06}
and \citet{Maret07a}. The abundance profiles were computed using a
chemical network that includes gas grain interactions, as well as the
fractionation reactions for carbon, oxygen, and deuterium
\citep{Bergin97}. These abundance profiles were used to compute the
line emission using a 1-D Monte Carlo radiative transfer code
\citep{Ashby00}. These model predictions were convolved at the
appropriate spatial resolution in order to be compared with the
observations, assuming a distance of 125 pc.

The density profile from \citet{Alves01} and the gas and dust
temperature from \cite{Bergin06b} were used. We have adopted the same
initial abundances and cosmic ionization rate as in the
\citet{Maret07a} ``best-fit'' model. As a first approach, we have
assumed that no systematic radial motions are present in the core. We
have used the turbulent velocity profile determined by
\cite{Bergin06b} from C$^{18}$O (1-0) line observations. The turbulent
velocity contribution is 0.3 $\mathrm{km \ s^{-1}}$ (FWHM) at the
surface of the core, and decreases towards the center down to 0.15 {km
  s$^{-1}$}. Rotation of the cloud has been neglected. Both the
$^{13}$CO (2-1) and the H$^{13}$CO$^{+}$ (1-0) lines have an
unresolved hyperfine structure that need to be taken into account in
order to properly model the line profiles. Following the approach used
by \citet{Tafalla06}, we have artificially broadened these two lines
by 0.13 km s$^{-1}$, which corresponds to the separation between the
two hyperfine components of each species \citep{Schmid-Burgk04}.

\begin{figure}
  \epsscale{0.75}
  \plotone{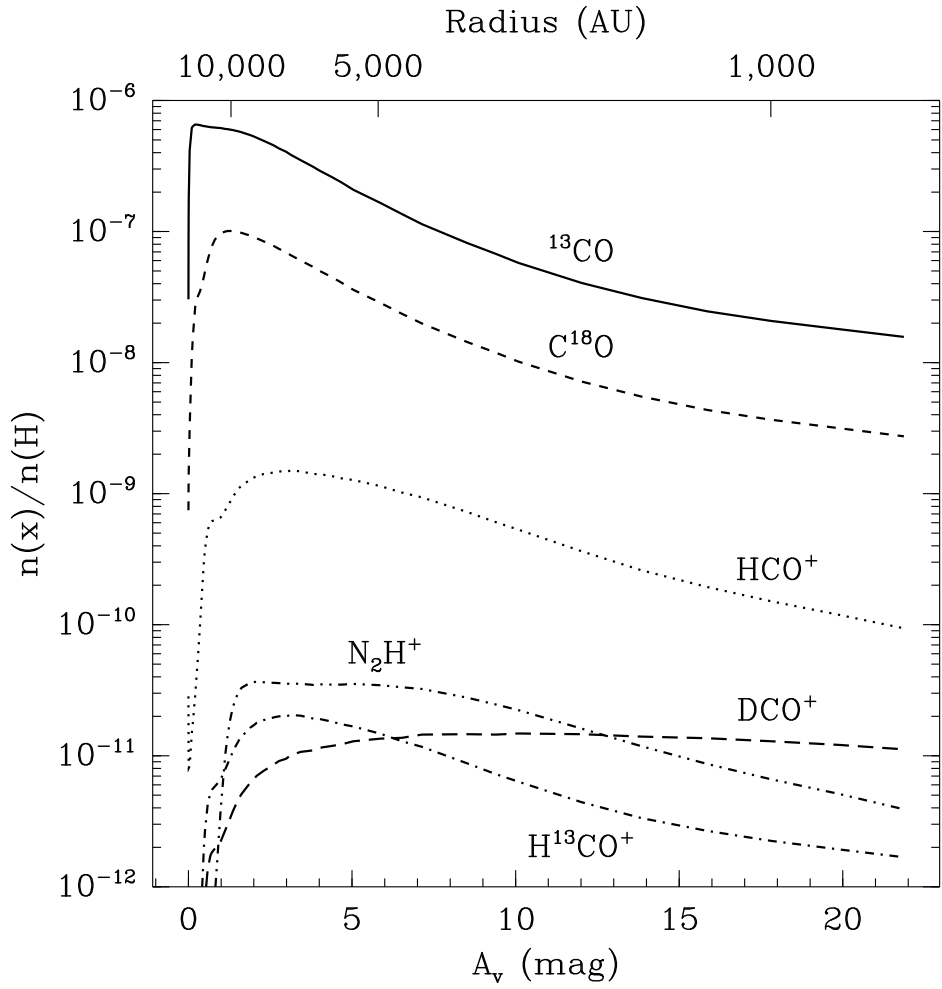}
  \caption{Abundances predicted by our chemical model for the observed
    species as a function of the visual extinction in the
    core.\label{fig2}}
\end{figure}

On Fig. \ref{fig2}, we present the abundances predicted by our
chemical model. From this figure we can roughly estimate which lines
originate in different parts of the core. For example, the HCO$^{+}$
abundance peaks in the outer part of the core ($A_{v} \sim 2$) and
decreases towards the center because of the freeze-out of its parent
molecule CO. Consequently, HCO$^{+}$ lines probe predominantly the
outer layer. Both N$_{2}$H$^{+}$ and DCO$^{+}$ abundances peak deeper
inside, at an $A_{v}$ of 5 and 13 respectively. Therefore lines from
these two species preferentially trace the inner regions of the
core. Of course, excitation effects are also important, and different
transitions of the same molecule are expected to probe different
depths in the core as well. For example, the HCO$^{+}$ (4-3) opacity
is likely lower that the (1-0) and (3-2) lines, and thus the HCO$^{+}$
(4-3) line probably arises from deeper regions than the HCO$^{+}$
(3-2) and (1-0) lines. Altogether, our observations suggest the
presence of complex radial motions in B68, with inwards motions in the
outer part of the core -- as indicated by the blue HCO$^{+}$ (1-0) and
(3-2) line profiles -- and outward motions in the inner region -- as
shown by the HCO$^{+}$ (4-3), DCO$^{+}$ (3-2) and N$_{2}$H$^{+}$ (1-0)
lines. In addition, the DCO$^{+}$ (2-1) line hints at the presence of
an intermediate static region between these two parts.

\begin{figure}
  \epsscale{0.75}
  \plotone{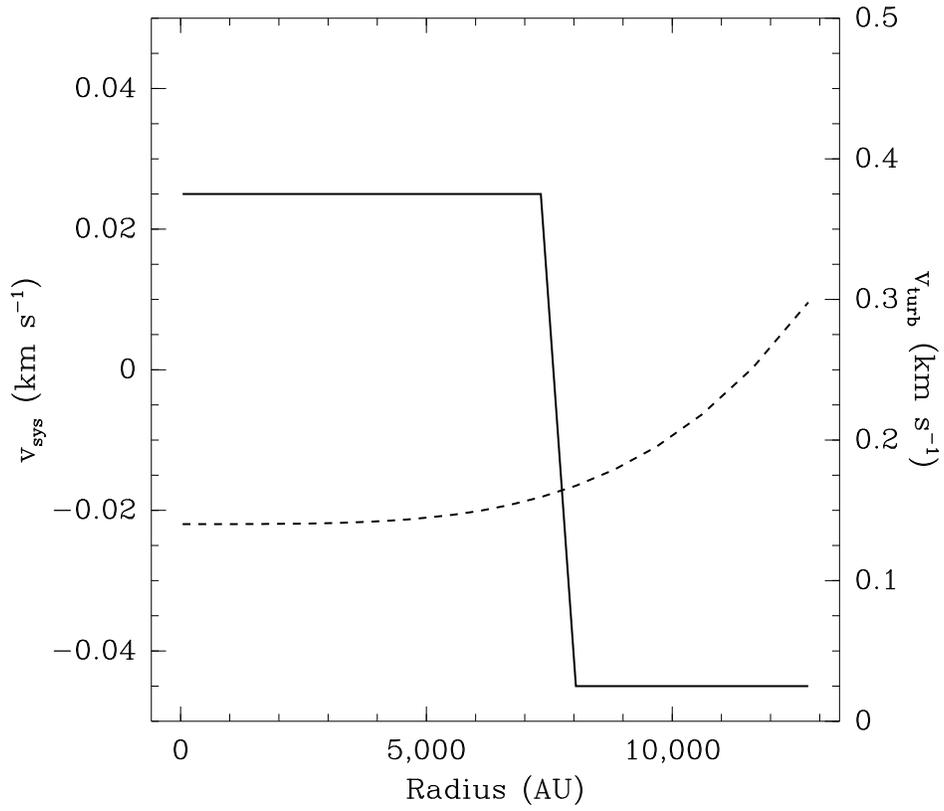}
  \caption{Best fit radial velocity profile. The solid line shows the
    systemic velocity profile (values are on the left axis). The
    dashed line shows the turbulent velocity profile \citep[values are
    on the right axis, from][]{Bergin06b}.\label{fig3}}
\end{figure}

In order to test this hypothesis, we have computed the line profiles
for the step-like velocity profile represented on Fig. \ref{fig3}. By
convention, negative velocities correspond to inward motions, while
positive velocities correspond to outward motions. The velocity is
assumed to be negative (-0.045 km s$^{-1}$) at radius greater than
8,000 AU, and positive (0.025 km s$^{-1}$) at smaller
radii. Fig. \ref{fig1} compares the observations with our model
predictions for this velocity profile. The fit is not
perfect. However, given the complexity of our modeling -- chemical
network coupled with a radiative transfer model -- the overall
agreement between the model predictions and observations is fairly
good. The model reproduces quite well the infall asymmetry of
HCO$^{+}$ (1-0) and (3-2), as well as the blue asymmetry of the (4-3)
line. H$^{13}$CO$^{+}$ (1-0), N$_{2}$H$^{+}$ (3-2), C$^{18}$O (1-0)
and $^{13}$CO (2-1) line profiles are also well matched.  Although the
model predicts correct integrated intensities for DCO$^{+}$ (3-2) and
N$_{2}$H$^{+}$ (1-0), it does not reproduce the blue asymmetries seen
in these lines. The predicted line profiles have a Gaussian shape,
whereas the observed lines are self-absorbed. In these cases, it
appears that our model underestimates the opacity of these lines.

\section{Discussion and conclusions}
\label{sec:discussion}

Our observations and modelling indicate that the outer part of the
core is collapsing while the inner part is expanding. The velocities
in both regions are relatively small (a few tens of meters per second)
and are largely sub-sonic. The transition between the collapsing and
expanding region occurs at a radius of about 8,000 AU. It is important
to note that the fit to the observations is probably not unique.
However, large changes in the transition and velocities between the
regions are likely ruled out. As already mentioned, our model does not
reproduce the asymmetries of the N$_{2}$H$^{+}$ (1-0) and DCO$^{+}$
line profiles, due to an underestimating of the opacity. There are
several possible explanations for this discrepancy. First, our
chemical model might not predict the correct abundance profile for
these two species. For example, increasing the DCO$^{+}$ abundance at
intermediate radii, with a subsequent reduction at small radii to keep
the column density constant would increase the opacity of line without
changing the line flux significantly (which is correctly
predicted). Indeed, our model predicts a DCO$^{+}$ (2-1) line
intensity at $10 < A_{v} < 20$ that is lower than observed \citep[see
Fig. 3 in][]{Maret07a}. Second, the physical structure of the core is
also somewhat uncertain. For example, the temperature profile may be
slightly different than the one predicted by \cite{Bergin06b}
\citep[see][]{Crapsi07}, although small changes ($\pm 3$ K) in the
temperature profile were found to have little effect on the line
intensities.  Turbulence at the center of the core might also be
different that what assumed here.  The turbulent profile was
determined by \cite{Bergin06b} using C$^{18}$O observations, which are
not a good probe of the innermost regions of the core because of heavy
depletion.  Decreasing the turbulent velocity profile in the innermost
region of the core would also increase the opacity of these two lines.
Regardless of the discrepancy with our model predictions, we emphasize
these two line profiles, together with the HCO$^{+}$ (4-3) line
profile, unambiguously indicate the presence of outwards motions at
the center of the core. We found that no constant infalling velocity
profile could reproduce the observations.

The step-like velocity profile we obtain is suggestive of the presence
of radial oscillations in the core. Small non-radial oscillations of
the outer layers of the core around a stable equilibrium have been
proposed by \citet{Lada03} and \cite{Redman06} to interpret the
asymmetry of the CS (2-1) and HCO$^{+}$ (3-2) lines observed in
Barnard 68. \cite{Lada03} also suggested that, in addition to the
surface oscillations seen in the CS (2-1) line, some radial
oscillation might also be present. Our observations and modeling are
in agreement with this hypothesis.

\cite{Keto05} proposed a hydrodynamic model for the evolution of cores
in quasi-equilibrium. In their modeling, a small perturbation of a
stable core (according to the Bonnor-Ebert criterion) is found to
result in damped oscillations around the equilibrium
position. \cite{Keto06} showed that the surface pattern observed in
the CS (2-1) line towards B68 is consistent with the (n = 1, l = m =
2) quadripolar oscillation mode for a specific orientation ($\lambda =
\theta = 30^\circ$). For the same oscillation mode, the radial
velocity is qualitatively similar to the one obtained in this study,
but is out of phase: the inner part of the core has a negative radial
velocity, while the outer part of the core has a positive velocity. At
the other half of the oscillation cycle, the velocity would be
reversed and in qualitative agreement with the profile we obtain
(E. Keto, \emph{priv. comm.})

As pointed out by \cite{Keto06} and \cite{Redman06}, the presence of
oscillations in B68 suggests that the core is relatively old,
typically older than a few sound speed crossing times ($\sim 10^{6}$
yr). This is comparable to the ambipolar diffusion timescale at the
center of the core and more than one order of magnitude higher than
the free-fall time scale \citep[$10^{6}$ and $7 \times 10^{4}$ yr
respectively;][]{Maret07a}. This is also about an order of magnitude
higher than the age of the core as determined from the degree of CO
depletion \citep[$10^{5}$yr;][]{Bergin02a,Bergin06b}\footnote{Strictly
  speaking, the age determined by this method is a lower limit of the
  real age of the cloud, because CO is assumed to be pre-existing, and
  the density is supposed to be constant \citep{Bergin06b}.}. For $t =
3 \times 10^{5}$ yr, \cite{Bergin06b} model predicts a C$^{18}$O (1-0)
line intensity that is three time smaller than the observations, while
at $t = 10^{6}$ yr, the greater CO depletion leads to an even greater
mismatch between observations and model predictions. The
N$_{2}$H$^{+}$, HCO$^{+}$ H$^{13}$CO$^{+}$, and DCO$^{+}$ would also
be underestimated by the model because of the freeze-out of their
parent of their precursors, N$_{2}$, CO and $^{13}$CO. Thus the
``chemical age'' of B68 inferred from the CO depletion appears to be
hard to reconcile with the ``dynamical age'' implied by the presence
of surface oscillations. It is also interesting to compare the sound
crossing time in B68 to typical statistical starless core lifetimes
\cite[see][for a review]{WardThompson07}. Using sub-millimeter
continuum emission maps, \cite{Kirk05} obtained lifetimes of $1-3
\times 10^{5}$ years which are in good agreement with the age of the
B68 determined from CO depletion, but again about an order of
magnitude higher than the sound crossing time.

The present study emphasizes the importance of understanding the
chemistry to constrain the kinematics of starless using spectral line
profiles. Because of the strong chemical gradients that exist in these
cores -- as well as excitation effects -- different lines can be used
to selectively trace different parts of the core. We have demonstrated
that, using lines that are appropriately chosen, it is possible to
reconstruct the core radial velocity profile. In B68, our observations
and modelling suggest the presence of complex line-of-sight motions
that are consistent radial oscillations. The presence of such
oscillation indicates that some core are long-lived, with lifetimes
about an order of magnitude higher than those derived from their
chemical evolution or sub-millimeter surveys. So far, oscillations
have been inferred in only one other core, FeSt 1-457
\citep{Aguti07}. Clearly, similar studies on a larger number of cores
are needed to establish if such cores are common.

\acknowledgments S.~M. wishes to thank Eric Keto for fruitful
discussions. The authors are also grateful to the referee for useful
and constructive comments. This work is supported by the National
Science Foundation under grant 0335207.

{\it Facilities:} \facility{IRAM:30m}, \facility{CSO}, \facility{APEX}


\clearpage

\end{document}